\documentclass[pdflatex,sn-nature]{sn-jnl}

\usepackage{graphicx}%
\usepackage{multirow}%
\usepackage{amsmath,amssymb,amsfonts}%
\usepackage{amsthm}%
\usepackage{mathrsfs}%
\usepackage[title]{appendix}%
\usepackage{xcolor}%
\usepackage{textcomp}%
\usepackage{manyfoot}%
\usepackage{booktabs}%
\usepackage{algorithm}%
\usepackage{algorithmicx}%
\usepackage{algpseudocode}%
\usepackage{listings}%
\usepackage{bm}%

\begin{document}

\title[PySlice]{PySlice: Routine Vibrational Electron Energy Loss Spectroscopy Prediction with Universal Interatomic Potentials}


\author*[1,2]{\fnm{Harrison A.} \sur{Walker}}\email{harrison.a.walker@vanderbilt.edu}

\author[3,4]{\fnm{Thomas W.} \sur{Pfeifer}}\email{pfeifertw@ornl.gov}

\author[5,7]{\fnm{Paul M.} \sur{Zeiger}}\email{pzeiger@uw.edu}

\author[3]{\fnm{Jordan A.} \sur{Hachtel}}\email{hachtelja@ornl.gov}

\author[1,2,6]{\fnm{Sokrates T.} \sur{Pantelides}}\email{pantelides@vanderbilt.edu}

\author*[3]{\fnm{Eric R.} \sur{Hoglund}}\email{hoglunder@ornl.gov}

\affil[1]{\orgdiv{Interdisciplinary Materials Science Program}, \orgname{Vanderbilt University}, \orgaddress{\city{Nashville}, \state{TN}, \country{USA}}}

\affil[2]{\orgdiv{Department of Physics and Astronomy}, \orgname{Vanderbilt University}, \orgaddress{\city{Nashville}, \state{TN}, \country{USA}}}

\affil[3]{\orgdiv{Center for Nanophase Materials Sciences}, \orgname{Oak Ridge National Laboratory}, \orgaddress{\city{Oak Ridge}, \state{TN}, \country{USA}}}

\affil[4]{\orgdiv{Department of Mechanical and Aerospace Engineering}, \orgname{University of Virginia}, \orgaddress{\city{Charlottesville}, \state{VA}, \country{USA}}}

\affil[5]{\orgdiv{Department of Materials Science and Engineering}, \orgname{University of Washington}, \orgaddress{\city{Seattle}, \state{WA}, \country{USA}}}

\affil[6]{\orgdiv{Department of Electrical and Computer Engineering}, \orgname{Vanderbilt University}, \orgaddress{\city{Nashville}, \state{TN}, \country{USA}}}

\affil[7]{\orgdiv{Department of Physics and Astronomy}, \orgname{Uppsala University}, \orgaddress{\city{Uppsala}, \country{Sweden}}}


\abstract{Vibrational spectroscopy in the electron microscope can reveal phonon excitations with nanometer spatial resolution, yet routine prediction remains out of reach due to fragmented workflows requiring specialized expertise. Here we introduce PySlice, the first publicly available implementation of the Time Autocorrelation of Auxiliary Wavefunction (TACAW) method, providing an automated framework that produces momentum- and energy-resolved vibrational electron energy loss spectra directly from atomic structures. By integrating universal machine learning interatomic potentials with TACAW, PySlice eliminates the bottleneck of per-system potential development. Users input atomic structures and obtain phonon dispersions, spectral diffraction patterns, and spectrum images through a unified workflow spanning molecular dynamics, GPU-accelerated electron scattering, and frequency-domain analysis. We outline the formulation behind the code, demonstrate its application to canonical systems in materials science, and discuss its use for advanced analysis and materials exploration. The modular Python architecture additionally supports conventional electron microscopy simulations, providing a general-purpose platform for imaging and diffraction calculations. PySlice makes vibrational spectroscopy prediction routine rather than specialized, enabling computational screening for experimental design, systematic exploration of phonon physics across materials families, and high-throughput generation of simulated data for training of future machine learning models.}


\maketitle

\section{Introduction}\label{sec1}
Vibrational electron energy loss spectroscopy (EELS) in the transmission electron microscope occupies a unique position among phonon measurement techniques, offering tunable access to both spatial and momentum resolution\cite{krivanek2014vibrational}. Optical methods such as Raman and infrared spectroscopy, including near-field variants like nano-FTIR and SNOM that achieve nanoscale spatial resolution, remain restricted to zone-center phonons ($\mathbf{q} \approx 0$) due to the negligible momentum of photons at vibrational energies. Inelastic neutron scattering and inelastic hard-X-ray scattering provide full momentum-resolved phonon dispersions, but require millimeter-scale samples and lack spatial resolution.

Vibrational EELS bridges this gap: by varying the probe convergence angle, the technique can trade between nanometer spatial resolution (focused beam) and precise momentum resolution (parallel beam), sampling the uncertainty relation $\Delta x \cdot \Delta p \gtrsim \hbar/2$ according to experimental requirements. This tunability enables investigations ranging from momentum-resolved phonon dispersions to spatially-resolved mapping of interfacial phonons, localized strain fields, and defect-induced vibrational states\cite{senga2019,hage2020science,hoglund2022nature,haas2023nanolettgb,hoglund2023advmat,xu2023natmat,hoglund2024advmat,yan2024nanolett}. The complexity of these experiments often necessitates predictive simulation to interpret results accurately and guide experimental design.

Several computational approaches address different aspects of vibrational EELS simulation. General-purpose multislice simulation codes such as abTEM\cite{madsen2021abtem} and Dr.\ Probe\cite{barthel2018dr} provide GPU-accelerated electron scattering calculations with sophisticated treatment of aberrations and detector geometries. These codes support frozen-phonon methods\cite{loane1991thermal} that approximate thermal disorder through static atomic displacements sampled from appropriate distributions. They capture diffuse scattering phenomena such as Kikuchi bands and thermal diffuse backgrounds. Separately, perturbative treatments based on harmonic phonon modes\cite{nicholls2019theory,senga2019,rossi2024qeels} provide theoretical models for the inelastic scattering cross-section, enabling prediction of EELS intensities from specific phonon excitations. An alternative approach by Dwyer\cite{dwyer2014localized} propagates electrons through mode-specific perturbed potentials, incorporating dynamical diffraction while still treating individual harmonic phonon modes.

Beyond these harmonic treatments, molecular dynamics (MD) simulations capture the full anharmonic dynamics of atomic motion. Methods such as those developed by Fransson et al. and Pfeifer et al.\cite{fransson2021dynasor,pfeifer2025phonon} extract dynamical structure factors directly from MD trajectories, providing efficient access to anharmonic-phonon dispersions that can be combined with theoretical cross-section formulations to predict EELS intensities. The Frequency Resolved Frozen Phonon Multislice (FRFPMS) method\cite{zeiger2020frfpms} and the Time Autocorrelation of Auxiliary Wavefunctions (TACAW)\cite{castellanos2025tacaw} combine beam propagation with MD trajectories, incorporating probe geometry and dynamical diffraction effects while accessing anharmonic dynamics.

FRFPMS performs multislice calculations on frequency-filtered-trajectories, such that computational cost scales with both the number of frequency bins and the trajectory duration. TACAW simplifies this workflow by moving the frequency decomposition from the input to the output: molecular dynamics generates a time series of atomic configurations sampling all thermally populated modes simultaneously, multislice propagation yields a time-dependent exit wavefunction, and a single temporal Fourier transform of the wavefunction data recovers the full frequency spectrum.

Recent developments in machine learning interatomic potentials (MLIPs) have expanded the scope of atomistic simulation. MLIPs are neural networks trained on databases of density-functional-theory (DFT) calculations that learn to predict energies and forces at near-DFT accuracy but orders of magnitude faster. Universal MLIPs (uMLIPs) such as ORB\cite{orb_ref}, MACE\cite{batatia2022mace}, CHGNet\cite{deng2023chgnet}, Allegro-FM\cite{allegrofm2025}, and UMA\cite{uma2025}, trained on diverse materials datasets encompassing much of the periodic table (e.g., the Materials Project\cite{jain2013materialsproject}, Alexandria\cite{alexandria2018}, OMat24\cite{omat24}, OMol25\cite{omol25}), provide transferable predictions of interatomic forces across chemical spaces without requiring per-system retraining. A user can supply any crystal structure and obtain physically reasonable dynamics without developing a custom potential. Concurrently, modern tensor libraries such as PyTorch\cite{pytorch} offer hardware abstraction across CPU and GPU backends and efficient batch operations that simplify implementation of numerical algorithms, enabling both the MLIPs and downstream multislice calculations to exploit GPU parallelism.

While each of these approaches addresses part of the vibrational EELS prediction problem, no existing framework integrates them into a unified, automated workflow. Frozen phonon methods lack the frequency resolution central to vibrational spectroscopy. Perturbation-theory approaches do not capture anharmonic effects. Existing multislice codes provide high-quality scattering simulations but lack integrated pipelines for vibrational EELS. TACAW offers the necessary physics, but practical implementation has required coordinating multiple specialized software packages---an MD engine, a multislice code, and custom analysis scripts---with no publicly available implementation. This fragmentation has kept vibrational EELS prediction a specialized undertaking requiring extensive custom development. The convergence of uMLIPs, GPU-accelerated multislice, and the TACAW formalism now creates an opportunity to unify these elements into a general-purpose prediction pipeline.

Here we present PySlice, the first publicly available implementation of the TACAW method. This modular Python framework integrates three key capabilities: (1) a unified interface to modern interatomic potentials including uMLIPs, with automated equilibration detection and MD trajectory management through ASE\cite{ase} and OVITO\cite{ovito}; (2) GPU-accelerated multislice calculations using PyTorch\cite{pytorch}, with persistent caching for simulation restarts and post-processing analyses; and (3) a structured data model for TACAW analysis that handles the time-to-frequency transformation and provides methods for extracting phonon dispersions, momentum-resolved spectra, and spectrum images. The object-oriented architecture maintains clean separation between MD, electron scattering, and spectroscopic analysis while enabling flexible composition of these elements for diverse simulation workflows.

We demonstrate PySlice through systematic application to transition metal dichalcogenides (TMDCs), where a single script produces material-specific phonon dispersions and spectral diffraction patterns across MoS$_2$, WS$_2$, WSe$_2$, and MoSe$_2$. We further demonstrate with vibrational spectrum images at a bulk Si/Ge heterostructure interface and a silicon substitution defect in graphene. Beyond TACAW simulations, PySlice's modular design supports conventional multislice imaging and diffraction simulations, providing a unified platform for TEM/STEM simulation tasks. By removing the practical barriers to vibrational EELS prediction and establishing consistent workflows, this work aims to accelerate both computational exploration and experimental validation of phonon spectroscopy in complex materials while fostering reproducibility in the community.

\section{Methods}\label{sec2}

\begin{figure*}[t]
\centering
\includegraphics[width=\textwidth,trim=0 100 0 100,clip]{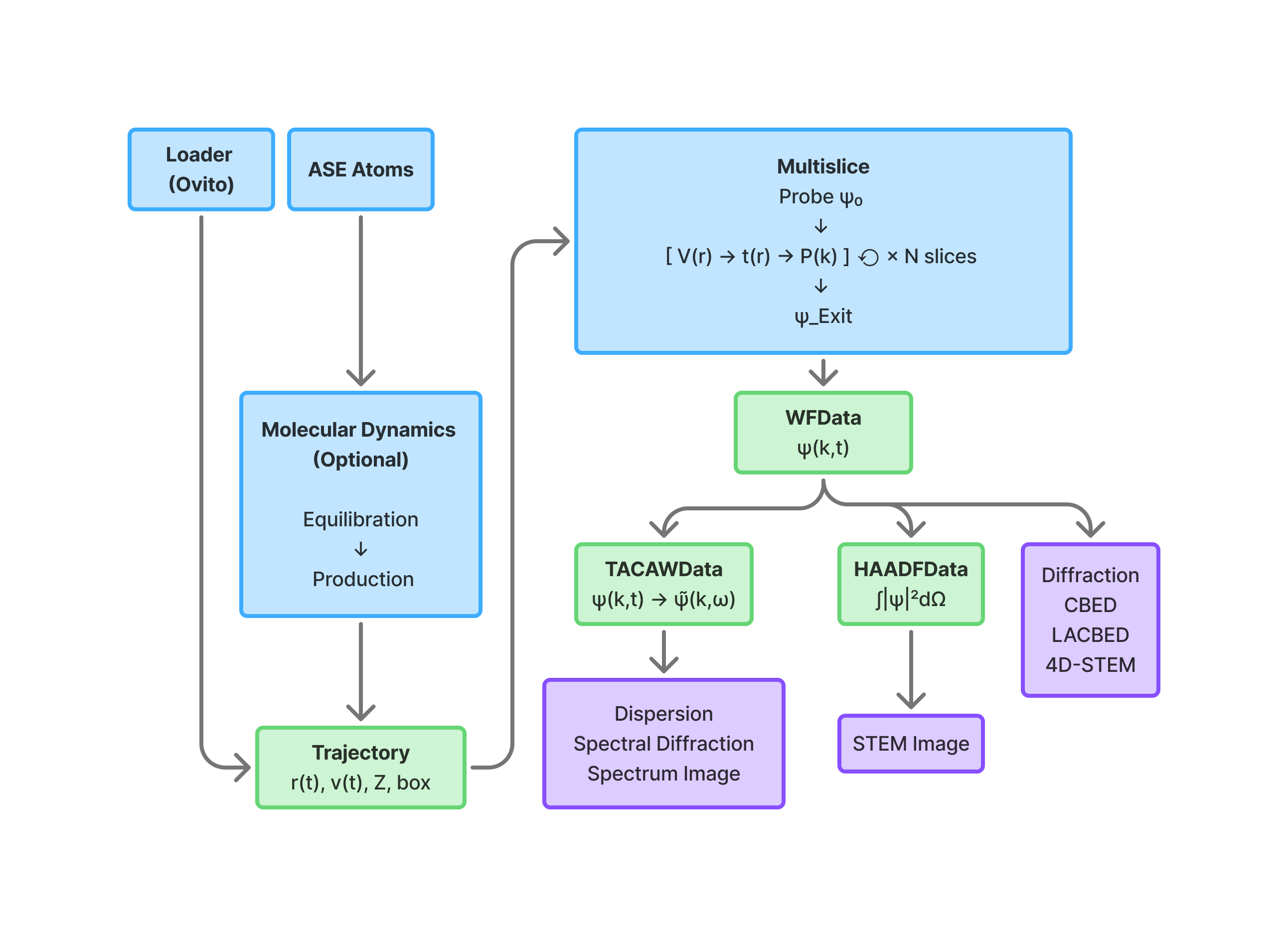}
\caption{\textbf{PySlice workflow for automated vibrational EELS prediction.} The pipeline integrates four stages: (1) input structures from CIF files, XYZ files, LAMMPS trajectories, or ASE Atoms objects; (2) optional molecular dynamics using universal machine learning interatomic potentials (ORB) with automated equilibration (NVT) and production (NVE) phases; (3) GPU-accelerated multislice simulation computing exit wavefunctions for each MD timestep; (4) TACAW analysis transforming time-domain wavefunctions to frequency-domain spectral intensities or conventional TEM/STEM such as frozen-phonon HAADF, TEM diffraction, or 4D-STEM. Data containers (\texttt{Trajectory}, \texttt{WFData}, \texttt{TACAWData}, \texttt{HAADFData}) manage information flow between stages.}
\label{fig:pipeline}
\end{figure*}

\subsection{Workflow Architecture}

Simulating vibrational EELS requires bridging three distinct computational domains: atomic dynamics, electron scattering, and spectroscopic analysis. PySlice addresses this through a modular pipeline (Fig.~\ref{fig:pipeline}) that cleanly separates these concerns while maintaining efficient data flow between stages.

The workflow begins with molecular dynamics simulations to generate thermally-populated atomic configurations. Rather than implementing yet another MD engine, PySlice's \texttt{MDCalculator} acts as a unified interface to modern interatomic potentials through ASE's molecular dynamics engine, with particular emphasis on the universal machine learning potential ORB. The \texttt{MDCalculator} handles equilibration detection automatically through convergence criteria such as temperature bounds and running averages, ensuring trajectories capture thermal phonon populations without manual intervention. For static imaging simulations, the MD is optional and trajectory will support only a single frame.

These time-resolved atomic configurations feed into the \texttt{MultisliceCalculator}, which must calculate a complete multislice simulation for each MD timestep. For a typical vibrational EELS simulation of several hundred timesteps at each probe position, this represents a substantial computational investment. Each timestep requires propagating the electron wavefunction slice-by-slice through the entire specimen thickness, with the atomic positions from that particular snapshot determining the scattering potential distribution.

The implementation uses PyTorch for GPU acceleration, with all numerically intensive operations executed on GPU hardware when available. The code automatically selects between CUDA, Apple Metal (MPS), or CPU backends depending on available hardware. Multislice calculations are cached at the frame level: each computed exit wave is written to disk as a NumPy array. This enables simulation restarts after interruption and allows post-processing analyses to operate on pre-computed wavefunction data without re-running the multislice calculations. While GPU acceleration is an integral feature to PySlice, the code also retains full functionality without PyTorch, optionally utilizing NumPy\cite{numpy} as the backend.

The \texttt{TACAWData} class then transforms the time-domain wavefunction data into frequency-domain spectra. For a trajectory with timesteps $\{t_i\}$ separated by $\Delta t$, the exit wavefunction $\psi(\mathbf{k},t_i)$ at each momentum point $\mathbf{k}$ is transformed: $\tilde{\psi}(\mathbf{k},\omega) = \mathcal{F}_t[\psi(\mathbf{k},t) - \langle\psi(\mathbf{k},t)\rangle_t]$. The mean subtraction isolates the time-varying component of the scattering, which contains the phonon information. The vibrational EELS intensity $I(\mathbf{k},\omega) = |\tilde{\psi}(\mathbf{k},\omega)|^2$ then reveals phonon excitations as peaks at energies $\hbar\omega$ and momenta $\hbar\mathbf{k}$ satisfying the phonon dispersion relation. The underlying physics is straightforward: as atoms vibrate at their characteristic phonon frequencies, they impart a periodic modulation (both temporally and spatially) on electron scattering potential. This modulation imparts the same temporal signature to the transmitted electron wavefunction, and elastic scattering to the corresponding wavevector is observed.
A Fourier transform in time extracts these frequency components, revealing phonon modes as features in the momentum- and energy-resolved intensity $I(\mathbf{k},\omega)$. PySlice organizes this analysis into a data structure with integrated methods for common phonon spectroscopy tasks: extracting dispersions along reciprocal space paths, generating momentum-resolved spectra, and computing spectrum images.

\subsection{Key Technical Elements}

\textbf{Multislice Propagation.} The multislice algorithm solves the electron scattering problem by decomposing the specimen into thin slices perpendicular to the beam direction. This approximation, valid when slice thickness is small compared to the transverse coherence length, allows the full 3D scattering problem to be treated as a sequence of 2D phase shifts (transmission) alternating with free-space propagation. Within each slice of thickness $\Delta z$, the wavefunction experiences a phase shift determined by the projected potential: $t(\mathbf{r}) = \exp(i\sigma V(\mathbf{r}))$, where $\sigma = 2\pi m_e \lambda / h^2$ is the interaction parameter and $m_e$ includes relativistic corrections. Between slices, Fresnel propagation in reciprocal space applies the operator $P(\mathbf{k}) = \exp(-i\pi\lambda\Delta z|\mathbf{k}|^2)$. The complete propagation through $N$ slices follows: $\psi_{\text{exit}} = P_{N-1} t_N \cdots P_1 t_2 P_0 t_1 \psi_{\text{probe}}$.

\textbf{Atomic Scattering Potentials.} Accurate atomic potentials are essential for quantitative EELS simulation. PySlice uses Kirkland's parameterized atomic structure factors\cite{kirkland2020advanced}, which represent decades of refinement against both quantum mechanical calculations and experimental measurements. For each element $Z$, the momentum-space form factor takes the form $f(\mathbf{q}) = \sum_{i=1}^{3} \left[\frac{a_i}{|\mathbf{q}|^2 + b_i} + c_i \exp(-d_i|\mathbf{q}|^2)\right]$, where the coefficients $\{a_i, b_i, c_i, d_i\}$ are tabulated. This parameterization captures both the high-angle scattering (first term) and low-angle scattering (second term) with sufficient accuracy for typical TEM conditions. The projected potential for each slice is assembled in reciprocal space by summing structure factors with appropriate phase factors for atomic positions: $V(\mathbf{k}) = \sum_j f_{Z_j}(\mathbf{k}) \exp(-2\pi i \mathbf{k} \cdot \mathbf{r}_j)$, where the sum runs over atoms within the slice z-bounds. An inverse FFT then yields the real-space projected potential.

\textbf{Molecular Dynamics.}
The TACAW method extracts phonon frequencies by Fourier transforming the time-dependent electron wavefunction, which inherits its temporal structure from atomic motion. This places two requirements on the MD trajectory: configurations must sample the thermal phonon population at the target temperature, and the time evolution must reflect deterministic phonon dynamics rather than thermostat-driven fluctuations. PySlice's \texttt{MDCalculator} addresses these requirements through a two-phase protocol: Langevin dynamics (NVT) drives the system to thermal equilibrium, after which velocity-Verlet integration (NVE) generates the production trajectory. Alternatively, production can use Langevin dynamics with very low friction, which maintains better energy conservation with some MLIPs while minimally perturbing the phonon dynamics.
Equilibration cannot be inferred from instantaneous temperature reaching its target as a system may pass through this value transiently while still relaxing toward thermal equilibrium. The \texttt{MDCalculator} instead evaluates convergence through three criteria assessed over sliding windows. First, the instantaneous temperature must have reached the target at least once, confirming sufficient thermal excitation. Second, the mean temperature over recent steps must remain within a tolerance of the target while its standard deviation stays below a stability threshold, ensuring the system fluctuates about the correct thermal state rather than drifting. Third, the relative standard deviation of potential energy per atom must fall below a threshold, indicating the system has settled into a stable region of configuration space rather than undergoing structural relaxation. Only when all three conditions are jointly satisfied does the simulation advance to production, automating what traditionally required manual inspection of thermalization curves. While the default parameters will work for most materials, the user has full control over the convergence criteria, timestep, and thermostat friction, which may be tuned depending on the task at hand. Guidance on selection of these parameters is discussed in supplemental materials. The resulting configurations and velocities from the MD run are automatically packaged into PySlice's \texttt{Trajectory} format with proper timestep metadata, enabling seamless handoff to the multislice calculation stage.

Spectral statistics can be further improved by averaging results over multiple independent trajectories or trajectory segments, though this multiplies the computational cost of the multislice calculations; guidance on this approach is provided in the supplemental materials.

\textbf{Data Structure Design.} The software architecture separates data storage from analysis through four core classes. The \texttt{Trajectory} class encapsulates atomic configurations and provides methods for accessing positions, velocities, and derived quantities (mean positions, displacements) without exposing the underlying storage format. The \texttt{WFData} class stores multislice results as complex wavefunctions indexed by probe position, time, and reciprocal space coordinates. This organization matches the natural axes of the TACAW calculation and enables efficient slicing operations for analysis. The \texttt{TACAWData} class inherits from \texttt{WFData} but replaces the time axis with frequency, providing specialized methods for phonon analysis: \texttt{dispersion()} for extracting dispersion relations along k-space paths, \texttt{spectral\_diffraction()} for momentum-resolved intensity at specific frequencies, and \texttt{spectrum\_image()} for position-resolved vibrational mapping in STEM geometries. Common TEM and 4D-STEM analysis can be performed using the \texttt{WFData} class, e.g. real-space and reciprocal-space plotting for imaging and diffraction. The \texttt{HAADFData} class is also made available to streamline the post-processing on the $x$,$y$,$k_x$,$k_y$ dataset contained in \texttt{WFData}, e.g. for conventional bright-field or annular dark field scanning probe imaging.

\section{Results}\label{sec3}

\begin{figure*}[t]
\centering
\includegraphics[width=\textwidth]{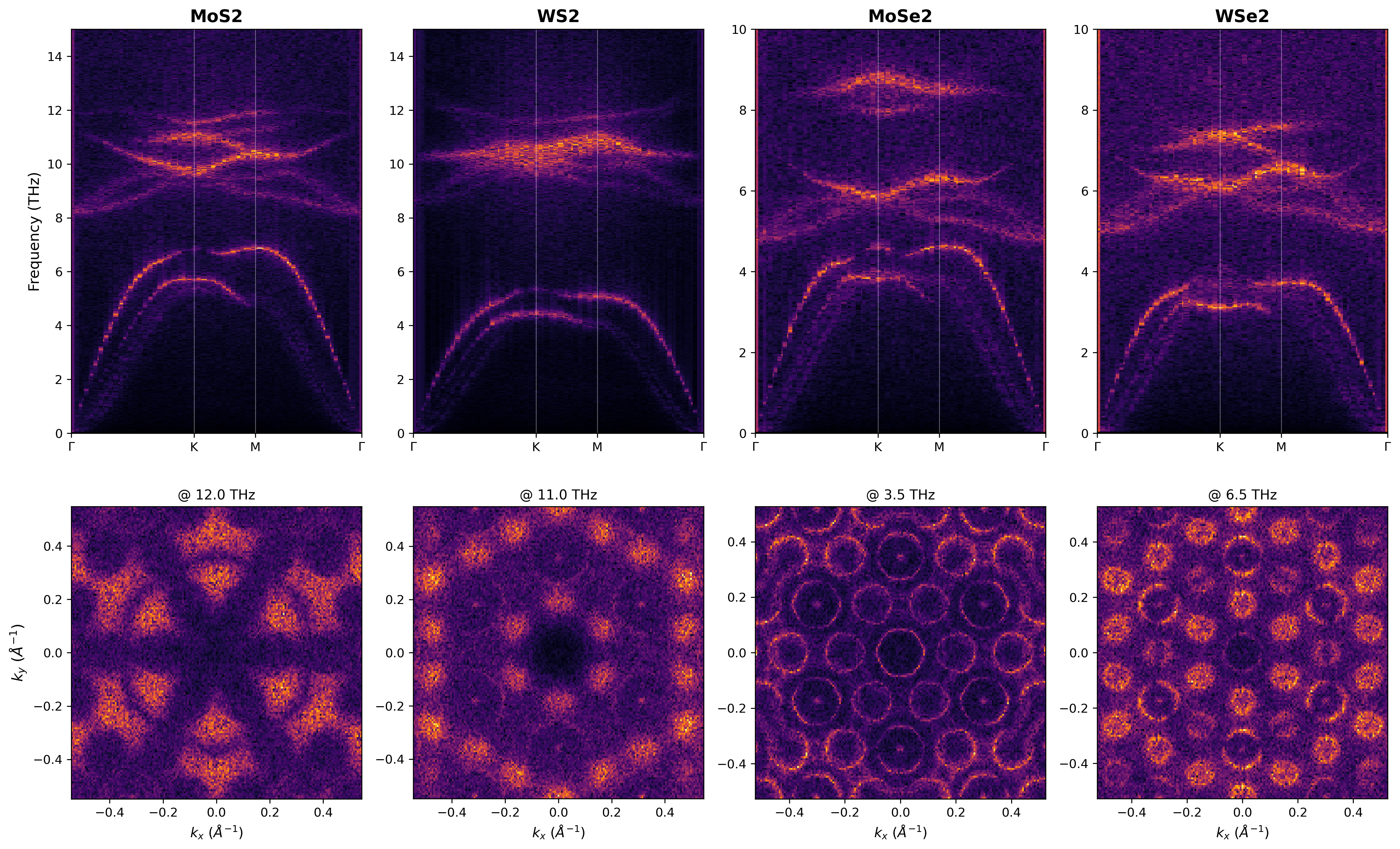}
\caption{\textbf{End-to-end automated TACAW spectra across the TMDC family.} All simulations were generated from a single PySlice script that handled the complete workflow: crystal structure generation with ASE, ORB-driven molecular dynamics with automatic equilibration, GPU-accelerated multislice propagation through hundreds of timesteps, and TACAW frequency-domain analysis. No intermediate file manipulation, format conversion, or manual parameter adjustment was required between materials. The resulting phonon dispersions and spectral diffraction patterns for MoS$_2$, WS$_2$, WSe$_2$ and MoSe$_2$ capture material-specific vibrational signatures arising from differences in atomic masses and bonding.}
\label{fig:tmdc_comparison}
\end{figure*}

\subsection{Automated Workflow Across TMDCs}

To validate the automated workflow, we applied PySlice to four monolayer transition metal dichalcogenides: MoS$_2$, WS$_2$, MoSe$_2$, and WSe$_2$ (Fig.~\ref{fig:tmdc_comparison}). All four materials were processed within a single script using identical simulation parameters, iterating over the chemical formulas. For each material, the workflow proceeded automatically from crystal structure generation through ORB-driven molecular dynamics, multislice propagation across MD frames, and TACAW frequency-domain analysis.

The resulting phonon dispersions (Fig.~\ref{fig:tmdc_comparison}, top row) show the expected material-specific features: acoustic branches rising from the $\Gamma$ point, optical branches at higher frequencies, and the characteristic flattening near zone boundaries. The sulfides (MoS$_2$, WS$_2$) exhibit optical modes extending to $\sim$14 THz, while the heavier selenides (MoSe$_2$, WSe$_2$) show correspondingly lower frequencies reaching $\sim$10 THz. Spectral diffraction patterns at representative frequencies (Fig.~\ref{fig:tmdc_comparison}, bottom row) display the hexagonal symmetry expected for these 2H-phase materials. These results demonstrate that a single automated pipeline produces physically reasonable, material-specific phonon spectra across a chemically related family. The fidelity of these predictions to experimental measurements depends on the underlying interatomic potential, which we discuss in Section~\ref{sec4}.

\subsection{Spatially-Resolved Phonon Modes}

\begin{figure*}[t]
\centering
\includegraphics[width=\textwidth]{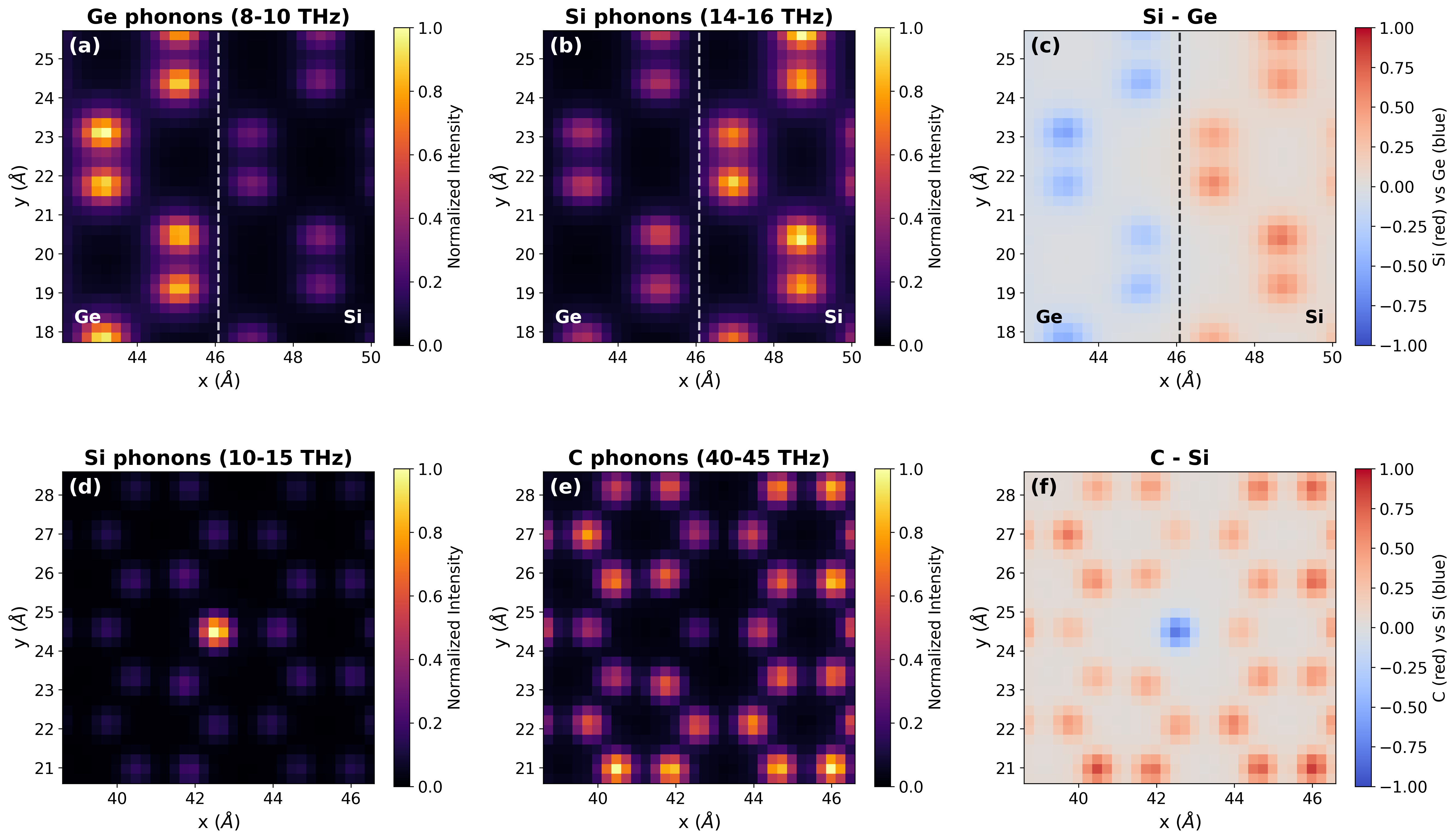}
\caption{\textbf{Spectrum Images of Localized Vibrations.} (a-c) Bulk Si/Ge heterostructure interface showing optical phonon confinement: (a) Ge optical modes ($\sim$8-10 THz) confined to the Ge region, (b) Si optical modes ($\sim$14-16 THz) confined to the Si region, and (c) the normalized difference of the two energy regions. (d-f) Silicon substitution defect in graphene showing localized defect modes: (d) Silicon vibrations (10-15 THz) localized to the isolated atom, while (e) Carbon vibrations (40-45 THz) appear throughout the structure. The difference plot in (f) highlights the Silicon impurity. These results demonstrate PySlice's capability to predict spatially-resolved vibrational signals at interfaces and point defects.}
\label{fig:spatial_localization}
\end{figure*}

PySlice's STEM scanning capability enables spatially-resolved phonon spectroscopy, revealing localized vibrational modes. We demonstrate this on two exemplary systems: a Si/Ge semiconductor interface and a silicon substitution in graphene (Fig.~\ref{fig:spatial_localization}).

The Si/Ge interface exemplifies phonon confinement at heterostructure boundaries. Due to the mass difference between Si (28 amu) and Ge (73 amu), optical phonon frequencies differ significantly between the two materials. TACAW spectrum imaging reveals that Ge optical modes (Fig.~\ref{fig:spatial_localization}a) are confined to the Ge region, while Si optical modes (Fig.~\ref{fig:spatial_localization}b) remain localized to the Si side of the interface. The normalized difference (Fig.~\ref{fig:spatial_localization}c) highlights this spatial separation. Acoustic modes, with wavelengths much longer than the interface width, propagate across both materials and are not spatially confined to either region in the spectrum images.

The silicon substitution in graphene demonstrates detection of point defect vibrations. The substantial mass difference between silicon (28 amu) and carbon (12 amu) produces distinct vibrational signatures: carbon atoms in the graphene lattice vibrate at characteristic frequencies near 40--45 THz, while the heavier silicon impurity exhibits lower-frequency modes in the 10--15 THz range. TACAW spectrum imaging at these frequencies reveals the silicon atom as a localized bright spot against the uniform graphene background, demonstrating PySlice's sensitivity to isolated atomic-scale defects.

\section{Discussion}\label{sec4}

\subsection{Implications for Materials Exploration}

Universal machine learning interatomic potentials make MD-based vibrational EELS prediction broadly accessible. Where researchers previously needed material-specific interatomic potentials, universal MLIPs provide transferable forces across chemical space. PySlice leverages this to enable automated spectroscopic predictions.

This capability opens several avenues for materials exploration. High-throughput screening becomes feasible: one can systematically compute vibrational spectra across materials families, alloy compositions, or defect configurations to identify systems with targeted phonon properties. The automated workflow also enables prediction-driven experimental design, where simulated spectra guide the selection of materials and measurement conditions before committing microscope time. Furthermore, the ability to generate large datasets of simulated spectra will provide training data for future machine learning models that analyze experimental EELS data.

The accuracy of predicted phonon frequencies and intensities depends on the underlying interatomic potential; PySlice faithfully propagates whatever physics the uMLIP encodes, but cannot correct for its deficiencies. Recent systematic benchmarking of uMLIPs against DFT phonon databases and experimental inelastic neutron scattering spectra demonstrates that models such as ORB achieve near-DFT accuracy for vibrational properties across diverse materials families\cite{han_cheng2025}. PySlice leverages this validated accuracy to enable automated vibrational EELS prediction. As uMLIPs continue to improve, the predictions from this workflow will correspondingly improve without requiring changes to the simulation infrastructure.

Computational parameters also constrain the accessible spectral range. Supercell size sets the minimum resolvable wavevector, while the MD timestep determines the maximum frequency via the Nyquist criterion. The total number of recorded frames controls frequency resolution. These are standard tradeoffs in any time-domain phonon simulation, and users can adjust supercell dimensions and trajectory length to match the spectral features of interest.

\subsection{Beyond Vibrational Spectroscopy}

\begin{figure}
    \centering
    \includegraphics[width=0.95\linewidth]{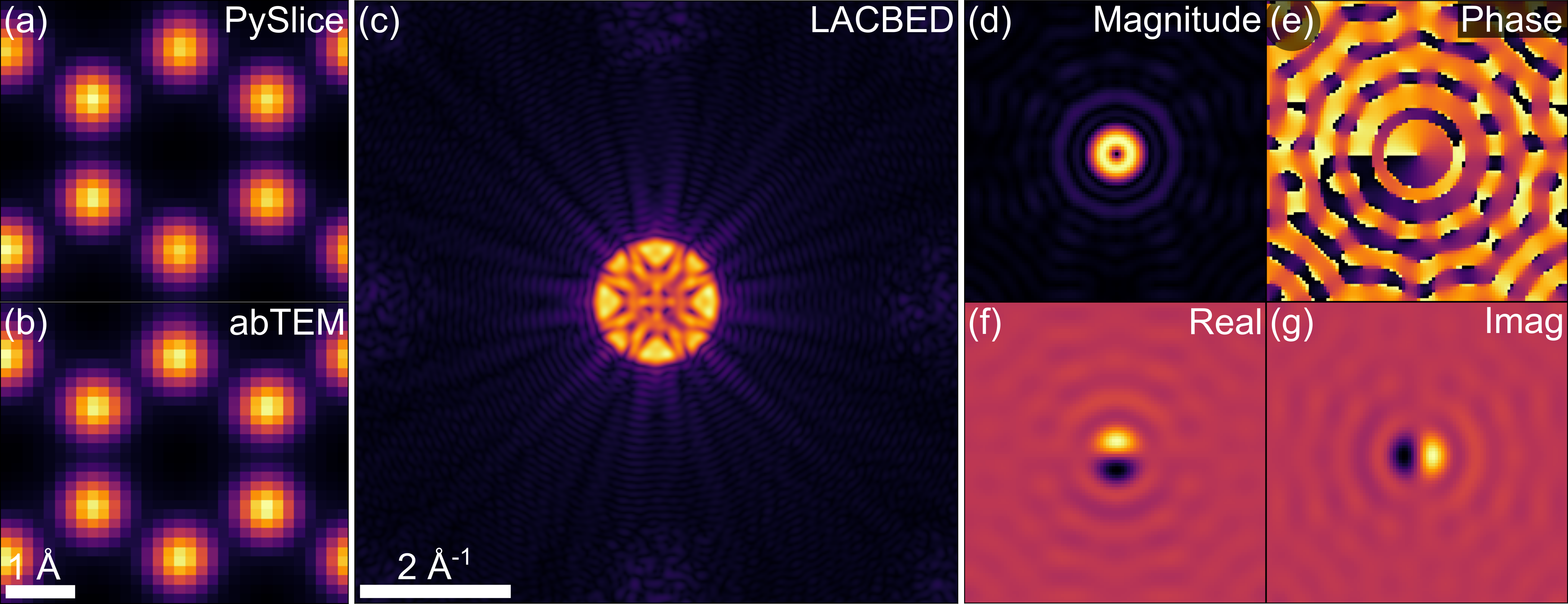}
    \caption{\textbf{Extended simulation capabilities beyond vibrational spectroscopy.} (a) Annular dark field (ADF) imaging validation against established multislice codes. (b) Large-angle convergent beam electron diffraction (LACBED) pattern demonstrating thickness-dependent dynamical diffraction effects. (c) Vortex beam probe with orbital angular momentum, illustrating PySlice's support for arbitrary wavefront profiles including chiral-sensitive illumination modes.}
    \label{fig:ExtendedOne}
\end{figure}

PySlice is architected around TACAW, but the same multislice machinery handles conventional STEM simulation as a natural byproduct (Fig.~\ref{fig:ExtendedOne}). Having both in one framework is convenient as users can validate probe configurations, generate comparison images, and run full vibrational analyses without switching tools.

For STEM imaging, the \texttt{HAADFData} class integrates scattered intensity within configurable collection angle ranges to produce annular dark-field images directly from multislice exit wavefunctions. High-angle configurations (inner angles of 50--80~mrad) yield Z-contrast images sensitive to atomic number, while adjusting the detector geometry accesses different scattering regimes. The underlying 4D dataset, diffraction patterns recorded at each probe position, remains accessible for more sophisticated 4D STEM analyses.

Realistic simulations often require modeling lens imperfections. PySlice implements the standard aberration function supporting defocus, astigmatism, coma, spherical aberration, and higher-order terms. Aberrations can be applied either during probe formation or post-hoc to stored wavefunctions, enabling efficient exploration of through-focal series or aberration correction strategies without repeating the multislice calculation.

The multislice propagation can optionally retain wavefunctions at each slice, providing depth-resolved scattering information relevant to convergent beam electron diffraction analysis. The \texttt{Probe} class also accepts arbitrary complex wavefront arrays, so users can construct custom phase profiles, spiral phases for vortex beams, for instance, and propagate them through the same infrastructure. This flexibility makes PySlice suitable for general TEM/STEM simulation tasks beyond its primary application in vibrational spectroscopy.

\section{Conclusion}\label{sec5}

We have presented PySlice, an automated framework for predicting vibrational electron energy loss spectra from atomic structures. By integrating universal machine learning interatomic potentials with GPU-accelerated multislice simulation and TACAW frequency-domain analysis, PySlice reduces the barrier to vibrational spectroscopy prediction from weeks of specialized development to hours of computation. We validated the approach on the transition metal dichalcogenide family, demonstrating material-specific phonon dispersions and spectral diffraction patterns from a single automated script. We further demonstrated vibrational spectrum images at heterostructure interfaces and point defects, capabilities that enable nanoscale phonon mapping in complex materials. PySlice provides a foundation for high-throughput spectroscopy prediction, training data generation, and prediction-driven experimental design in vibrational EELS. Example scripts and complete workflows are provided in the supplemental materials and code repository.

\backmatter

\bmhead{Acknowledgements}

H.A.W. \& S.T.P. were supported by the U.S. Department of Energy, Office of Basic Energy Sciences (DOE-BES), Division of Materials Sciences and Engineering grant DE-FG02-09ER46554, the U.S. Department of Energy, Office of Science User Facility, and by the McMinn Endowment.

H.A.W. \& S.T.P. acknowledge computing resources from the National Energy Research Scientific Computing Center (NERSC), a U.S. Department of Energy Office of Science User Facility located at Lawrence Berkeley National Laboratory, operated under Contract No. DE-AC02-05CH11231.

This material is based upon work supported by the National Science Foundation Graduate Research Fellowship Program under Grant No.\ 2444112. Any opinions, findings, and conclusions or recommendations expressed in this material are those of the author(s) and do not necessarily reflect the views of the National Science Foundation.

T.W.P. \& E.R.H. acknowledge the support of ORNL's Materials Characterization Core provided by UT-Battelle, LLC, under Contract No. DE-AC05-00OR22725 with the DOE and sponsored by the Laboratory Directed Research and Development Program of Oak Ridge National Laboratory, managed by UT-Battelle, LLC, for the U.S. Department of Energy.

This research used resources of the Compute and Data Environment for Science (CADES) at the Oak Ridge National Laboratory, which is supported by the Office of Science of the U.S. Department of Energy under Contract No. DE-AC05-00OR22725.

P.M.Z. acknowledges funding from the Swedish Research Council under grant no. 2024-06617.

Notice: This manuscript has been authored by UT-Battelle, LLC, under contract DE-AC05-00OR22725 with the US Department of Energy (DOE). The US government retains and the publisher, by accepting the article for publication, acknowledges that the US government retains a nonexclusive, paid-up, irrevocable, worldwide license to publish or reproduce the published form of this manuscript, or allow others to do so, for US government purposes. DOE will provide public access to these results of federally sponsored research in accordance with the DOE Public Access Plan (\url{https://www.energy.gov/doe-public-access-plan}).

\section*{Declarations}

\begin{itemize}
\item \textbf{Funding:} See Acknowledgements.
\item \textbf{Conflict of interest:} The authors declare no competing interests.
\item \textbf{Data availability:} All data presented in this work were generated using PySlice and can be reproduced using the scripts provided in the code repository.
\item \textbf{Code availability:} PySlice is available at \url{https://github.com/h-walk/PySlice}.
\item \textbf{Author contribution:} H.A.W. wrote the manuscript. H.A.W., T.W.P., and P.M.Z. developed the PySlice software. J.A.H., S.T.P., and E.R.H. provided guidance. All authors reviewed and edited the manuscript.
\end{itemize}

\bibliography{Refs}

\end{document}


\title[PySlice Supplemental]{Supplemental Material: PySlice---Routine Vibrational Electron Energy Loss Spectroscopy Prediction with Universal Interatomic Potentials}

\author*[1,2]{\fnm{Harrison A.} \sur{Walker}}\email{harrison.a.walker@vanderbilt.edu}

\author[3,4]{\fnm{Thomas W.} \sur{Pfeifer}}\email{pfeifertw@ornl.gov}

\author[5,7]{\fnm{Paul M.} \sur{Zeiger}}\email{pzeiger@uw.edu}

\author[3]{\fnm{Jordan A.} \sur{Hachtel}}\email{hachtelja@ornl.gov}

\author[1,2,6]{\fnm{Sokrates T.} \sur{Pantelides}}\email{pantelides@vanderbilt.edu}

\author*[3]{\fnm{Eric R.} \sur{Hoglund}}\email{hoglunder@ornl.gov}

\affil[1]{\orgdiv{Interdisciplinary Materials Science Program}, \orgname{Vanderbilt University}, \orgaddress{\city{Nashville}, \state{TN}, \country{USA}}}

\affil[2]{\orgdiv{Department of Physics and Astronomy}, \orgname{Vanderbilt University}, \orgaddress{\city{Nashville}, \state{TN}, \country{USA}}}

\affil[3]{\orgdiv{Center for Nanophase Materials Sciences}, \orgname{Oak Ridge National Laboratory}, \orgaddress{\city{Oak Ridge}, \state{TN}, \country{USA}}}

\affil[4]{\orgdiv{Department of Mechanical and Aerospace Engineering}, \orgname{University of Virginia}, \orgaddress{\city{Charlottesville}, \state{VA}, \country{USA}}}

\affil[5]{\orgdiv{Department of Materials Science and Engineering}, \orgname{University of Washington}, \orgaddress{\city{Seattle}, \state{WA}, \country{USA}}}

\affil[6]{\orgdiv{Department of Electrical and Computer Engineering}, \orgname{Vanderbilt University}, \orgaddress{\city{Nashville}, \state{TN}, \country{USA}}}

\affil[7]{\orgdiv{Department of Physics and Astronomy}, \orgname{Uppsala University}, \orgaddress{\city{Uppsala}, \country{Sweden}}}

\maketitle

\section{Supplemental Material}

\subsection*{Automated Molecular Dynamics Equilibration}

PySlice implements an automated two-phase molecular dynamics protocol: NVT equilibration followed by NVE (or low-friction NVT) production. The equilibration phase uses a Langevin thermostat with moderate friction to rapidly bring the system to thermal equilibrium. Once convergence is detected, the simulation switches to production dynamics with minimal thermostat interference, ensuring clean phonon dynamics suitable for TACAW analysis.

Convergence during equilibration is assessed using multiple criteria evaluated over sliding windows. The algorithm requires all of the following conditions to be satisfied simultaneously:

\begin{enumerate}
\item \textbf{Temperature stability}: The standard deviation of temperature over the most recent $N_\text{temp}$ steps must fall below a threshold:
\begin{equation}
\sigma_T = \sqrt{\langle T^2 \rangle - \langle T \rangle^2} < \sigma_T^\text{max}
\end{equation}

\item \textbf{Temperature on target}: The mean temperature must be within tolerance of the target:
\begin{equation}
|\langle T \rangle - T_\text{target}| < \Delta T_\text{tol}
\end{equation}

\item \textbf{Energy stability}: The relative standard deviation of potential energy must fall below a threshold:
\begin{equation}
\frac{\sigma_E}{|\langle E \rangle|} < \epsilon_E
\end{equation}
\end{enumerate}

A critical feature of the algorithm is the \texttt{reached\_target} flag: the system must first reach the target temperature at least once before convergence can be declared. This prevents false convergence during the initial heating phase when temperature and energy may appear stable but at incorrect values. A minimum step requirement ensures sufficient statistics are accumulated before convergence checks begin.

Table~\ref{tab:convergence} lists the default convergence parameters with their descriptions.

\begin{table}[h]
\caption{Default parameters for MD equilibration convergence checking.}
\label{tab:convergence}
\small
\begin{tabular}{lll}
\toprule
Parameter & Default & Description \\
\midrule
\texttt{temp\_threshold} & 5.0 K & Max allowed $\sigma_T$ \\
\texttt{temp\_tolerance} & 20.0 K & Max $|{\langle T \rangle - T_\text{target}}|$ \\
\texttt{energy\_threshold} & 0.01 & Max rel.\ $\sigma_E/|\langle E \rangle|$ \\
\texttt{temp\_window} & 50 & Temp.\ averaging window \\
\texttt{energy\_window} & 50 & Energy averaging window \\
\texttt{min\_equil\_steps} & 100 & Min steps before checking \\
\texttt{max\_equil\_steps} & 20000 & Forced transition \\
\bottomrule
\end{tabular}
\end{table}

These defaults are intentionally permissive to ensure robust convergence across diverse materials without user intervention. For quantitative studies where precise temperature control is important, users should consider tightening \texttt{temp\_tolerance} (e.g., to 10~K) and increasing the averaging windows to better capture slow relaxation dynamics.

\begin{figure}[h]
\centering
\includegraphics[width=\columnwidth]{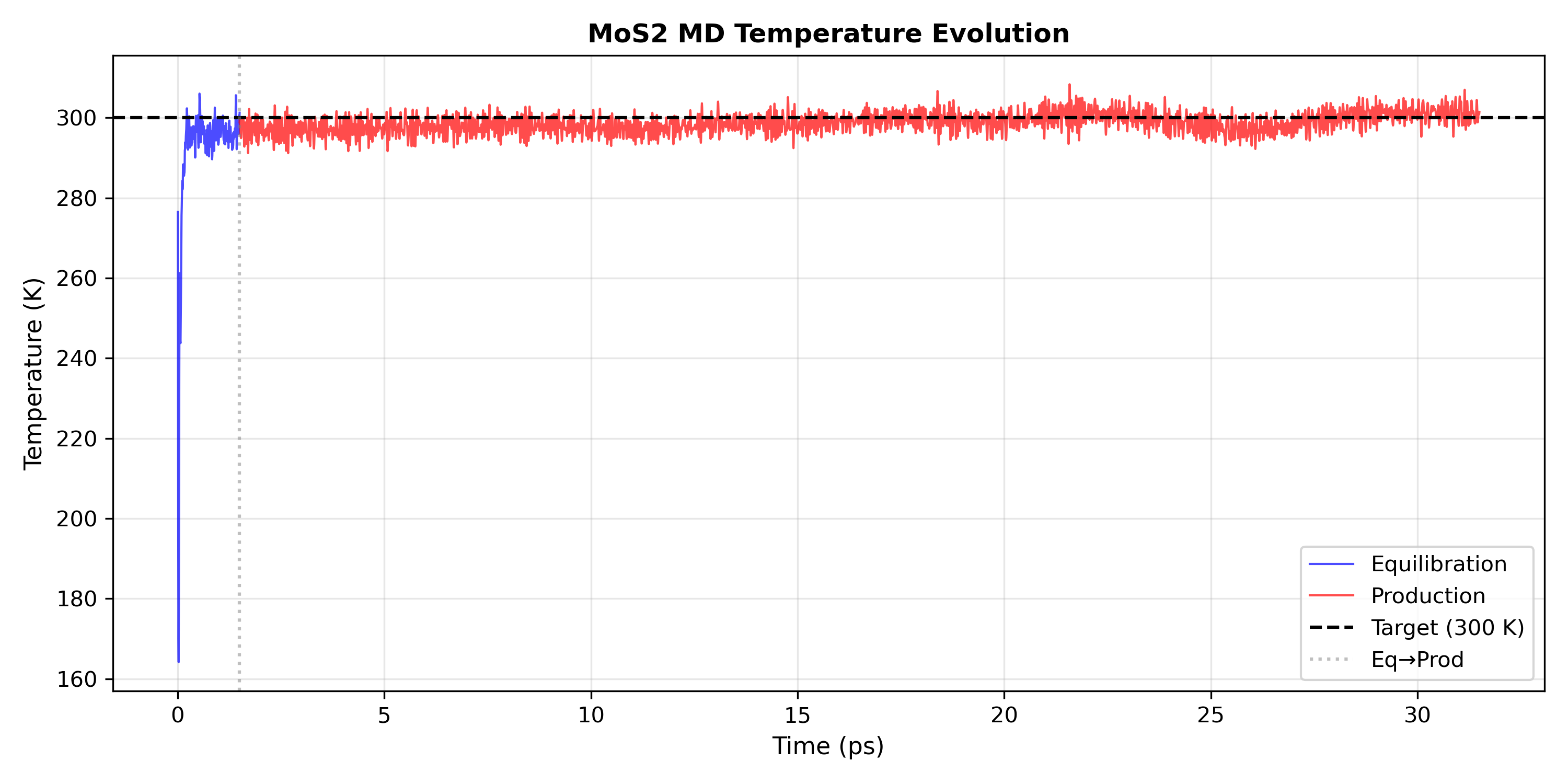}
\caption{MD equilibration for MoS$_2$ using the ORB potential. Temperature evolution during equilibration shows the system reaching thermal equilibrium, after which convergence criteria are satisfied and production dynamics begin.}
\label{fig:convergence}
\end{figure}

Figure~\ref{fig:convergence} shows a representative equilibration trajectory for MoS$_2$ at 300~K. The temperature initially fluctuates as the system equilibrates from its starting configuration, then stabilizes around the target. Once all convergence criteria are met, the simulation automatically transitions to the production phase.

\subsection*{Sampling Considerations}

Several sampling parameters determine the resolution and range of accessible phonon frequencies in TACAW analysis:

\textbf{Supercell size} determines the minimum resolvable wavevector. For a simulation cell with dimension $L$ along a given direction:
\begin{equation}
\Delta k = \frac{1}{L}
\end{equation}
Larger supercells provide finer sampling of the Brillouin zone but increase computational cost.

\textbf{Trajectory length} controls frequency resolution. For $N$ frames separated by time interval $\Delta t$:
\begin{equation}
\Delta f = \frac{1}{N \cdot \Delta t}
\end{equation}
Longer trajectories provide finer frequency resolution but require more storage and computation.

\textbf{Sampling timestep} sets the Nyquist limit on the maximum resolvable frequency:
\begin{equation}
f_\text{max} = \frac{1}{2\Delta t_\text{sample}}
\end{equation}
Note that the sampling timestep for TACAW analysis can differ from the MD integration timestep---one may integrate at 1~fs for numerical stability while saving frames at 10~fs intervals.

\textbf{Trajectory averaging} can improve signal-to-noise ratio. Multiple independent MD trajectories initialized with different random velocities sample different regions of phase space; averaging their TACAW spectra reduces statistical noise while preserving the underlying phonon structure.

\begin{figure*}[ht]
\centering
\includegraphics[width=\textwidth]{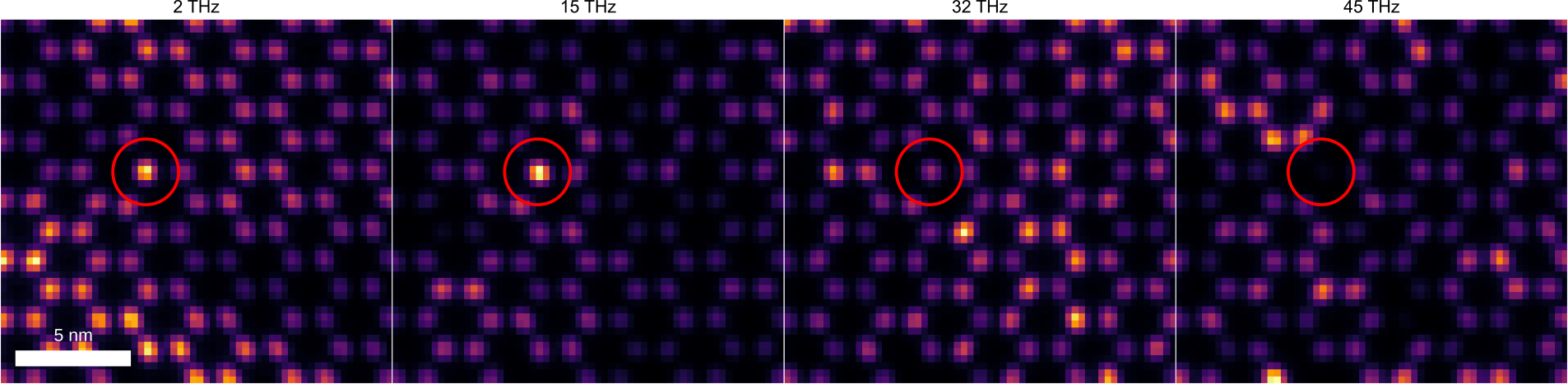}
\caption{Spectrum images of a silicon substitution defect in graphene at representative frequencies (2, 15, 32, and 45~THz) computed from a single trajectory. The red circle marks the silicon atom. Intensity variations among equivalent carbon sites demonstrate that insufficient trajectory length leads to non-ergodic sampling artifacts. Averaging over multiple independent trajectories or extending the simulation duration improves convergence.}
\label{fig:sampling}
\end{figure*}

Figure~\ref{fig:sampling} illustrates the consequences of insufficient sampling. Spectrum images of a silicon substitution in graphene computed from a single short trajectory show intensity variations at equivalent carbon sites, reflecting incomplete ergodic sampling of the phonon phase space. The red circle marks the silicon atom position. Longer trajectories or averaging over multiple independent runs improves convergence of the spectral statistics.

\subsection*{Physical Approximations and Limitations}

The multislice algorithm implemented in PySlice computes \textit{elastic} electron scattering through the time-dependent specimen potential. Several approximations are inherent in this approach:

\textbf{Elastic scattering formalism.} The transmission function at each slice applies a phase shift based on the instantaneous electrostatic potential:
\begin{equation}
T(\mathbf{r}_\perp) = \exp[i\sigma V(\mathbf{r}_\perp)]
\end{equation}
No explicit energy transfer between the electron and phonons is computed. Instead, phonon frequencies are extracted from the temporal modulation of diffraction intensity as atoms move along their MD trajectories. This approach is valid because phonon energies (meV scale) are negligible compared to the electron kinetic energy (100~keV), so the electron's wavelength and scattering geometry are effectively unchanged by the energy transfer.

\textbf{Decoupled electron-sample dynamics.} Atomic trajectories are computed independently of the electron beam. The simulation assumes:
\begin{itemize}
\item No electron-induced heating or radiation damage
\item No perturbation of atomic motion by the probing electron
\item The electron samples the thermal phonon population without influencing it
\end{itemize}
This approximation is valid for the low-dose conditions typical of vibrational EELS measurements, where beam-induced effects are minimized experimentally.

\subsection*{Extensibility to Other Atomistic Dynamics}

PySlice accepts any trajectory providing atomic positions as a function of time. The framework is not limited to thermal molecular dynamics---any atomistic dynamics method producing time-resolved positions will work with the TACAW analysis pipeline. Potential applications include:

\textbf{Atomistic spin dynamics}: For studying magnon-phonon coupling and magnetic excitations in materials where spin and lattice degrees of freedom are coupled.

\textbf{Path integral molecular dynamics}: To include quantum nuclear effects at low temperatures, where zero-point motion and tunneling become significant.

\textbf{Driven dynamics}: Non-equilibrium simulations such as laser excitation or applied fields could reveal transient phonon populations.

The key requirement is that the trajectory provides: (i) a positions array of shape $(N_\text{frames}, N_\text{atoms}, 3)$, (ii) atomic species identification, and (iii) the simulation cell geometry.

\subsection*{Minimal Working Example}

The following Python script demonstrates a complete TACAW workflow: generating a crystal structure, running molecular dynamics, computing multislice exit wavefunctions, and extracting phonon dispersions. This example computes the phonon dispersion of bulk silicon along the $\Gamma \to X$ direction.

\begin{lstlisting}
from ase.build import bulk
from pyslice import MDCalculator, \
    MultisliceCalculator, TACAWData
import numpy as np

# 1. Create structure: 20x20x5 Si supercell
atoms = bulk("Si", "diamond", a=5.431, cubic=True)
atoms = atoms * (20, 20, 5)

# 2. Run MD with automatic equilibration
md = MDCalculator(
    model_name="orb-v3-conservative-inf-omat")
md.setup(
    atoms=atoms,
    temperature=300,           # K
    timestep=2.0,              # fs
    production_steps=1000,     # 2 ps trajectory
    save_interval=5,           # Save every 10 fs
)
trajectory = md.run()
# 200 frames at 0.01 ps spacing -> Delta_f = 0.5 THz
# f_max = 50 THz (Nyquist limit)

# 3. Run multislice (parallel beam for TACAW)
calc = MultisliceCalculator()
calc.setup(
    trajectory,
    aperture=0,                # Parallel beam
    voltage_eV=100e3,          # 100 keV
    sampling=0.1,              # Angstrom/pixel
    slice_thickness=0.5,       # Angstrom
)
wf_data = calc.run()

# 4. Compute TACAW spectrum
tacaw = TACAWData(wf_data)

# 5. Extract dispersion along Gamma -> X
a = 5.431
kx = np.linspace(0, 1/(2*a), 100)  # Gamma to X
ky = np.zeros_like(kx)
dispersion = tacaw.dispersion(kx, ky)

# 6. Plot spectral diffraction at 15 THz
import matplotlib.pyplot as plt
Z = tacaw.spectral_diffraction(15.0)
plt.imshow(Z**0.25, cmap="inferno")
plt.savefig("spectral_15THz.png")
\end{lstlisting}

\textbf{Resolution analysis for this example:} The $20 \times 20 \times 5$ supercell with $a = 5.431$~\AA{} gives real-space extent $L = 108.6$~\AA{} and reciprocal-space resolution $\Delta k = 1/L \approx 0.009$~\AA$^{-1}$. The \texttt{sampling} parameter (0.1~\AA) sets the maximum wavevector: $k_\text{max} = 5$~\AA$^{-1}$, far exceeding the Si BZ boundary at $k_X \approx 0.092$~\AA$^{-1}$. Approximately 10 k-points sample the $\Gamma \to X$ path. The 200-frame trajectory at 0.01~ps spacing provides frequency resolution $\Delta f = 0.5$~THz and Nyquist limit $f_\text{max} = 50$~THz---sufficient for Si optical phonons ($\sim$15~THz).

For STEM spectrum imaging, add probe positions to the \texttt{MultisliceCalculator}:

\begin{lstlisting}
# Define probe scan region
xs = np.linspace(0, 20, 32)
ys = np.linspace(0, 20, 32)

calc.setup(
    trajectory,
    aperture=30,               # 30 mrad convergence
    voltage_eV=100e3,
    sampling=0.1,
    probe_xs=xs,               # Scan grid
    probe_ys=ys,
)
wf_data = calc.run()
tacaw = TACAWData(wf_data)

# Extract spectrum image at specific frequency
spectrum_image = tacaw.spectrum_image(15.0)
\end{lstlisting}

For loading existing trajectories (LAMMPS, ASE, XYZ):

\begin{lstlisting}
from pyslice import Loader

# LAMMPS dump file
traj = Loader(
    "dump.lammpstrj",
    timestep=0.01,             # ps
    atom_mapping={1: "Si", 2: "Ge"}
).load()

# ASE trajectory file
traj = Loader("production.traj", timestep=0.01).load()
\end{lstlisting}

Additional examples covering HAADF imaging, aberrations, LACBED patterns, and vortex beams are provided in the \texttt{tests/} directory of the repository.

\subsection*{Code Availability}

All code, including examples and test scripts, is available at \url{https://github.com/h-walk/PySlice}.